\documentclass[twocolumn,secnumarabic,amssymb, superscriptaddress, aps, prd, floatfix]{revtex4-1}
\usepackage{amsmath}
\usepackage{graphicx}
\usepackage{float}
\usepackage{tabularx}
\usepackage{color}
\usepackage{parskip}
\usepackage{xfrac}
\usepackage{lineno}
\begin{document}

\title{A method for the experimental measurement of bulk and shear loss angles in amorphous thin films}

\date{\today}

\author{Gabriele Vajente}\email{vajente@caltech.edu} \affiliation{California Institute of Technology, \\ LIGO Laboratory MC 100-36, \\ 1200 E. California Blvd. Pasadena (CA) USA}
\author{Mariana Fazio} \affiliation{Department of Physics, Colorado State University, Fort Collins, CO 80523, USA}
\author{Le Yang} \affiliation{Department of Chemistry, Colorado State University, Fort Collins, CO 80523, USA}
\author{Anchal Gupta} \affiliation{California Institute of Technology, \\ LIGO Laboratory MC 100-36, \\ 1200 E. California Blvd. Pasadena (CA) USA}
\author{Alena Ananyeva} \affiliation{California Institute of Technology, \\ LIGO Laboratory MC 100-36, \\ 1200 E. California Blvd. Pasadena (CA) USA}
\author{Garilynn Billinsley} \affiliation{California Institute of Technology, \\ LIGO Laboratory MC 100-36, \\ 1200 E. California Blvd. Pasadena (CA) USA}
\author{Carmen S. Menoni}  \affiliation{Department of Physics, Colorado State University, Fort Collins, CO 80523, USA} \affiliation{Department of Chemistry, Colorado State University, Fort Collins, CO 80523, USA}  

\begin{abstract}
Brownian thermal noise is a limiting factor for the sensitivity of many high precision metrology applications, among other gravitational-wave detectors. The origin of Brownian noise can be traced down to internal friction in the amorphous materials that are used for the high reflection coatings. To properly characterize the internal friction in an amorphous material, one needs to consider separately the bulk and shear losses. In most of previous works the two loss angles were considered equal, although without any first principle motivation. In this work we present a method that can be used to extract the material bulk and shear loss angles, based on current state-of-the-art coating ring-down measurement systems. We also show that for titania-doped tantala, a material commonly used in gravitational-wave detector coatings, the experimental data strongly favor a model with two different and distinct loss angles, over the simpler case of one single loss angle.
\end{abstract}

\maketitle

\section{Introduction}\label{sec:introduction}

High precision optical metrology relies on high finesse and low loss optical resonant cavities, built with high reflectivity dielectric mirrors. The ultimate limit to the length stability of such cavities is often determined by thermal motion of the cavity components. In many cases, such as in interferometric gravitational wave (GW) detectors \cite{aligo, avirgo, kagra, geo600}, the limit thermal noise comes from the Brownian motion of the dielectric coatings deposited on the mirrors \cite{thermal-noise}, and composed of alternating layers of amorphous oxides: silica and titania-doped tantala for the Advanced GW detectors \cite{coatings-gw}. The amplitude of Brownian noise can be linked to the material internal friction by use of the Fluctuation-Dissipation Theorem \cite{fdt, levin}. In the simplest possible approximation the energy lost per cycle due to internal friction is modeled as a fraction of the total elastic energy $E$ stored in one of the resonator eigenmodes, using one single number usually called the \emph{loss angle} $\phi$:
\begin{equation}\label{eq:loss-angle}
\left<\Delta E\right>_{\mathrm{cycle}} = \phi \left<E\right>
\end{equation}
If the surface of the mirror is probed with a Gaussian laser beam with beam radius $w$, then in the simple approximation described above the displacement noise due to Brownian motion has a power spectral density \cite{PSD} given by \cite{reid2016}
\begin{equation}\label{eq:thermal-noise}
S(f) = \frac{4 k_B T}{\pi^2 \, f} \frac{(1 + \nu_S)(1 - 2 \nu_S)}{Y_S} \, \frac{d}{w} \, \phi_C
\end{equation} 
where $f$ is the frequency, $k_B$ is Boltzmann's constant, $T$ the temperature, $Y_S$ and $\nu_S$ the Young's modulus and Poisson ratio of the mirror substrate, $d$ is the coating thickness and $\phi_C$ the coating average loss angle. In this model the beam is assumed to be much larger than the film thickness, and there is no distinction between energy lost in the shear and bulk deformations of the mirror.

However, even for an amorphous material, the bulk and shear moduli are not equal, and therefore by extension there is no reason to assume that the bulk and shear loss angles have the same value. The theory of room temperature loss in amorphous materials \cite{phillips72, topp96} ascribes the energy loss mechanism to the presence of two-level systems, effectively described as double-well potentials with thermally excited tunneling between the two minima. The material mechanical loss is determined by the density of the two-level systems, by the distribution of the potential wells and barriers, and by the coupling of the two-level systems to the macroscopic elastic strain. There is no reason to assume that the two-level systems would couple in the same way to bulk and shear strains. Lacking a theoretical or phenomenological reason to assume the contrary, in computing the thermal noise due to the elastic energy loss in a multilayer coating, one needs to take into account both shear and bulk deformations and allow for the loss mechanisms to be different. The resulting displacement noise depends on the value of both bulk and shear loss angles in a way more complex than what shown in equation \ref{eq:thermal-noise} \cite{hong2013}. In particular, it is generally believed that the shear loss angle is more relevant than the bulk loss angle, when the beam size is comparable with the film thickness. Therefore, to have an accurate estimate of the Brownian noise in an optical system, it is important to have a reliable measurement of both loss angles.

The most common technique to measure the loss angle(s) of a thin film is to deposit it on a high quality resonator, and measure the decay time $\tau$ of a subset of the eigenmodes. This can be accomplished by exciting the resonator and tracking the oscillation amplitude of each mode over time:
\begin{equation}
A_i(t) = A_0 e^{-t/\tau_i}
\end{equation}
Some excess energy loss is always present for all modes, due for example to contact at the suspension point or substrate clamp. It is generally possible to find a suitable set of eigenmodes for which recoil losses are negligible, and are well decoupled from the environment. Typically those modes allow probing the material loss angle over a sufficiently large range of frequencies. Measuring the decay time of this set of eigenmodes allows probing the value and frequency dependency of the loss angles. For each eigenmode at a frequency $f_i$, the decay time $\tau_i$ is linked to the coated resonator quality factor $Q_i$ and loss angle $\phi_i$ by the following relations
\begin{equation}
\phi_i = \frac{1}{Q_i} = \frac{1}{\pi f_i \tau_i}
\end{equation}
The loss angle $\phi_i$ of the coated resonator should not be confused with the loss angle of the materials. It is related to the total elastic energy loss per cycle, and we can therefore divide it in two terms: a contribution coming from the substrate $\phi_i^{(sub)}$ and a contribution coming from the thin film $\phi_i^{(film)}$. The contribution of each term to the total loss angle is weighted by the amount of elastic energy which is stored in the substrate and in the film, on average:
\begin{eqnarray}
\phi_i^{(\mathrm{coated})} &=& \frac{E_i^{(\mathrm{sub})} \phi_i^{(\mathrm{sub})} + E_i^{(\mathrm{film})} \phi_i^{(\mathrm{film})}}{E_i^{(\mathrm{sub})} + E_i^{(\mathrm{film})}}  \nonumber \\
&=& (1 - D_i)  \phi_i^{(\mathrm{sub})}  + D_i \phi_i^{(\mathrm{film})} 
\end{eqnarray}
where we have introduced the mode dependent \emph{dilution factor} $D_i=E_i^{(\mathrm{film})}/ E_i^{(\mathrm{tot})}$. The substrate loss angle can be measured before any film is deposited, and it is usually assumed to remain unchanged by the deposition process. Therefore the difference of loss angles as measured before and after the film is deposited can be used to extract the loss angle of the material composing the film. We define the \emph{excess loss} of the coated sample as
\begin{equation}\label{eq:excess-1}
\delta \phi_i = \phi_i^{(\mathrm{coated})} - (1- D_i) \phi_i^{(\mathrm{sub})} = D_i \phi_i^{(\mathrm{film})}
\end{equation}
The dilution factors $D_i$ can be computed using finite element simulations of the resonators, knowing the elastic properties of the material, or extracted directly from the change in the eigenmode resonant frequencies \cite{resonant}. Since we are interested in measuring the bulk and shear loss angles $\phi_{B,i}$ and $\phi_{S,i}$, we need to modify the model in equation \ref{eq:excess-1} above as follows
\begin{equation} \label{eq:excess-2}
\delta \phi_i = D_{B,i} \phi_{B,i} + D_{S, i} \phi_{S,i}
\end{equation}
where we defined the new bulk and shear dilution factors as $D_{B,i}  = E^{(\mathrm{film})}_{B,i}/E^{(\mathrm{tot})}_{i}$ and $D_{S,i}  = E^{(\mathrm{film})}_{S,i}/E^{(\mathrm{tot})}_{i}$, so that $D_i = D_{B,i} + D_{S,i}$. Below we will describe how the elastic properties can be extracted from the modal frequencies and then used to calculate the dilution factors using a finite element model.

In this paper we describe how it is possible to analyze the resonant mode decay times of a thin film deposited on a silica disk-shaped substrate measured in a Gentle Nodal Suspension \cite{gens_cesarini, gens_vajente}, and express the film properties in terms of bulk and shear loss angle. In summary the analysis proceed in several steps. First of all, the elastic properties of the film are extracted from the shift in the resonator eigenmodes due to the addition of the film. This estimate is carried out with a Bayesian inference analysis and includes uncertainties that model the limited knowledge and possible evolution with heat treatment of the film density and thickness. More details on this first step in section \ref{sec:measurements}. The posterior probability distribution of the elastic properties are then used as priors for another Bayesian inference analysis, where the measured excess losses introduced in equation \ref{eq:excess-1} or equation \ref{eq:excess-2} are estimated based on a model of the material loss angle(s). This procedure factor into the posterior distribution of the loss angle the uncertainties in the material properties and possible correlation between the model parameters. More details in section \ref{sec:analysis}.

Analysis of measurements in terms of bulk and shear loss angles were done in the past for films on a cantilever composed of alternating layers of silicon nitride and silica \cite{pan2018}, and for a titania-doped tantala film on a disk suspended with a fiber \cite{abernathy2018}. 

We show the result of our analysis for a titania-doped tantala film as an example, and discuss how the experimental data favor a model with different bulk and shear loss angle over a simpler model with equal loss angles. The material studied here is comparable to what was considered in \cite{abernathy2018}, and we note that the results we obtain are different from those obtained in the previous work. More on this topic in section \ref{sec:analysis}. Finally, in section \ref{sec:thermal-noise} we discuss how the measured loss angles impact the estimate of thermal noise for the Advanced LIGO gravitational wave detector.

\section{Measurements} \label{sec:measurements}

\begin{table*}[t]
\centering
\begin{tabular}{| l | c | c | c |}
\hline
 & \textbf{As deposited} & \textbf{Annealed 500$^\circ$C} & \textbf{Annealed 600$^\circ$C}  \\
 \hline
\textbf{Young's modulus} $Y$ [GPa]  &  118$\pm$3  &   120$\pm$3  &   128$\pm$4   \\
\textbf{Poisson ratio} $\nu$ &  0.396$\pm$0.016 &  0.407$\pm$0.013 &  0.346$\pm$0.019      \\
\hline
\textbf{Cation concentration} & \multicolumn{3}{c|}{73\% Ta, 27\% Ti } \\
\textbf{Thickness} $t$ [nm]& \multicolumn{3}{c|}{268 $\pm$ 13} \\
\textbf{Density} $\rho$ [kg/m$^3$]&  \multicolumn{3}{c|}{6640 $\pm$ 300 } \\
\hline
\end{tabular}
\caption{Measured and estimated parameters of the titania-doped tantala thin film studied in this work. The thickness was measured on the as deposited samples, and the density estimated from the composition. The film elastic properties come from fits to the resonant mode data, as explained in the text. The uncertainties in thickness and density account for possible variations upon annealing, as explained in the main text.\label{tab:parameters}}
\end{table*} 

The substrates used in this work consist of fused silica disks, 75 mm in diameter and 1 mm thick, supported at the center by a  \emph{gentle nodal suspension} \cite{gens_cesarini, gens_vajente}. All the disk eigenmodes that have null deformation at the disk center are accessible in this system, and have very low recoil losses ($Q^{(\mathrm{sub})} \gtrsim 10^8$). The largest fraction of elastic energy is stored in shear deformation, but depending on the mode shape, in particular on the number of radial nodes, there are non negligible amounts of energy in the bulk deformation, allowing us to disentangle the two contributions.

The gentle nodal suspension allows simultaneous tracking of all modes, providing a measurement of both the frequency and the decay time of each mode. All substrates are characterized prior to coating, to measure the substrate loss angles $\phi_i^{(\mathrm{sub})}$ and the frequency of each mode. A 270-nm-thick film of titania-doped tantala (27\% cation concentration of titania) was then deposited with ion beam sputtering on one face of the substrates. The coated samples were then measured again, to obtain a new set of mode frequencies and decay times. The samples were then subjected to a heat treatment (\emph{annealing}), consisting of a slow ramp up to a target temperature, hold for ten hours, and then a slow ramp down to room temperature. The samples measured for this work have been annealed at 500, 600 and 700$^\circ$C. The film annealed at 700$^\circ$C showed signs of micro-crystallization, and therefore the corresponding results are not considered in this work. Ring downs were measured after each heat treatment step, resulting in a set of excess loss angles $\left\{\delta \phi_i \right\}$ for the as-deposited samples and the annealed samples.

The film thickness $t$ was measured with ellipsometry, and the relative concentration of titania and tantala was estimated from the measured refractive index and X-ray photoelectron spectroscopy. The material density $\rho$ was estimated with a linear interpolation between the two oxide component densities, weighted with the measured oxide concentration.

The thin film changes the flexural rigidity of the disk, resulting in a shift of all resonant mode frequencies. The relative difference between the coated and uncoated disk frequencies is roughly constant between 1 and 30 KHz, and equal to about 300 ppm, with variation between modes of the order of 10-30 ppm, related to the film Poisson ratio. We used a finite element analysis (FEA) carried out in COMSOL to find the values of the film material Young's modulus $Y$ and Poisson ratio $\nu$ that best reproduce the measured changes in resonant frequencies \cite{coatings-gw}. Instead of using directly COMSOL in the fit procedure, we first produced a random sampling of the film properties space $[Y, \nu, t, \rho]$ and run a FEA for each point. We then fit a third order polynomial function of $Y, \nu, t$ and $\rho$ to the simulated frequency shifts, obtaining a fast semi-analytical model that is accurate within tens of mHz. Using this fast model, we carried out a Bayesian inference analysis \cite{bayesian} to estimate the probability distribution and the confidence intervals for $Y$ and $\nu$. Table \ref{tab:parameters} summarizes all the measured parameters of the thin films. The results are dependent on the thickness and density of the film. The reader unfamiliar with Bayesian inference analysis can refer for example to \cite{silvia2006, bayesian, bayesian2011} for an introduction. In section \ref{sec:analysis} we also describe the basics of Bayesian inference, focusing on the application to the extraction of bulk and shear loss angles from the measurements.

In this analysis we assumed that thickness and density are constant, since we do not have yet a measurement of how those film properties change with annealing. This assumption is likely wrong, since changes of density, thickness and refractive index have been observed for other amorphous materials \cite{coatings-gw, annealing-density1,annealing-density2}. However, we note that the estimate of $Y$ and $\nu$ depends mostly on the product of thickness and density, that is, the surface density of the material. Therefore, even though density and thickness could each vary, if the annealing does not cause any loss of material from the film, we expect that the product of density and thickness will remain constant and the estimate of the Young's modulus and Poisson ratio to be correct. Nevertheless, in the analysis we accounted for possible untracked changes by allowing a $\pm 5$\% uncertainty in the measured values for both thickness and density. 

Two samples were coated with nominally equal materials and deposition procedure. The two samples have been measured separately, and the results collated together in all computations.

\section{Loss angle analysis}\label{sec:analysis}

\begin{figure*}[tb] 
\begin{center}
\includegraphics[width=0.66\columnwidth]{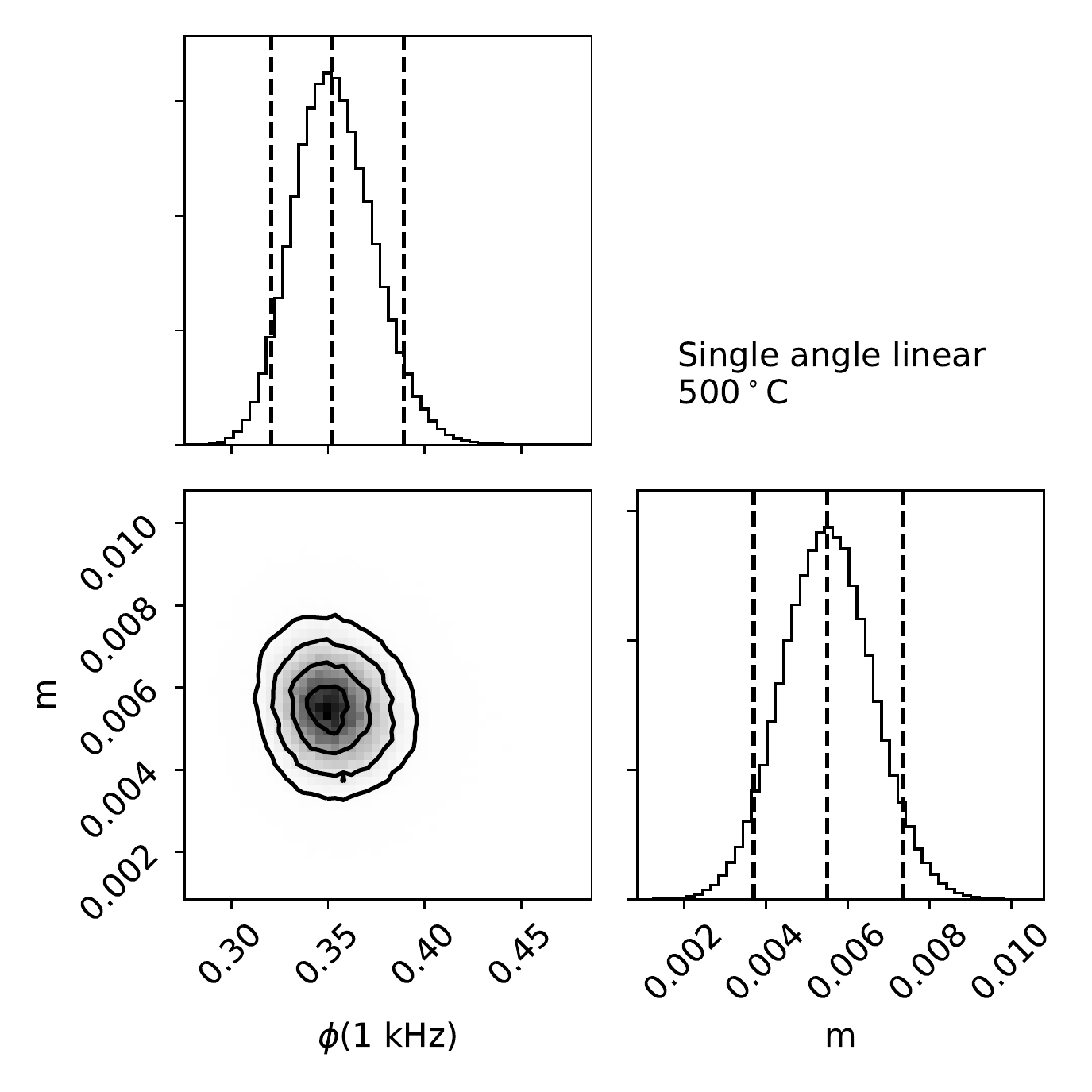} \includegraphics[width=1.2\columnwidth]{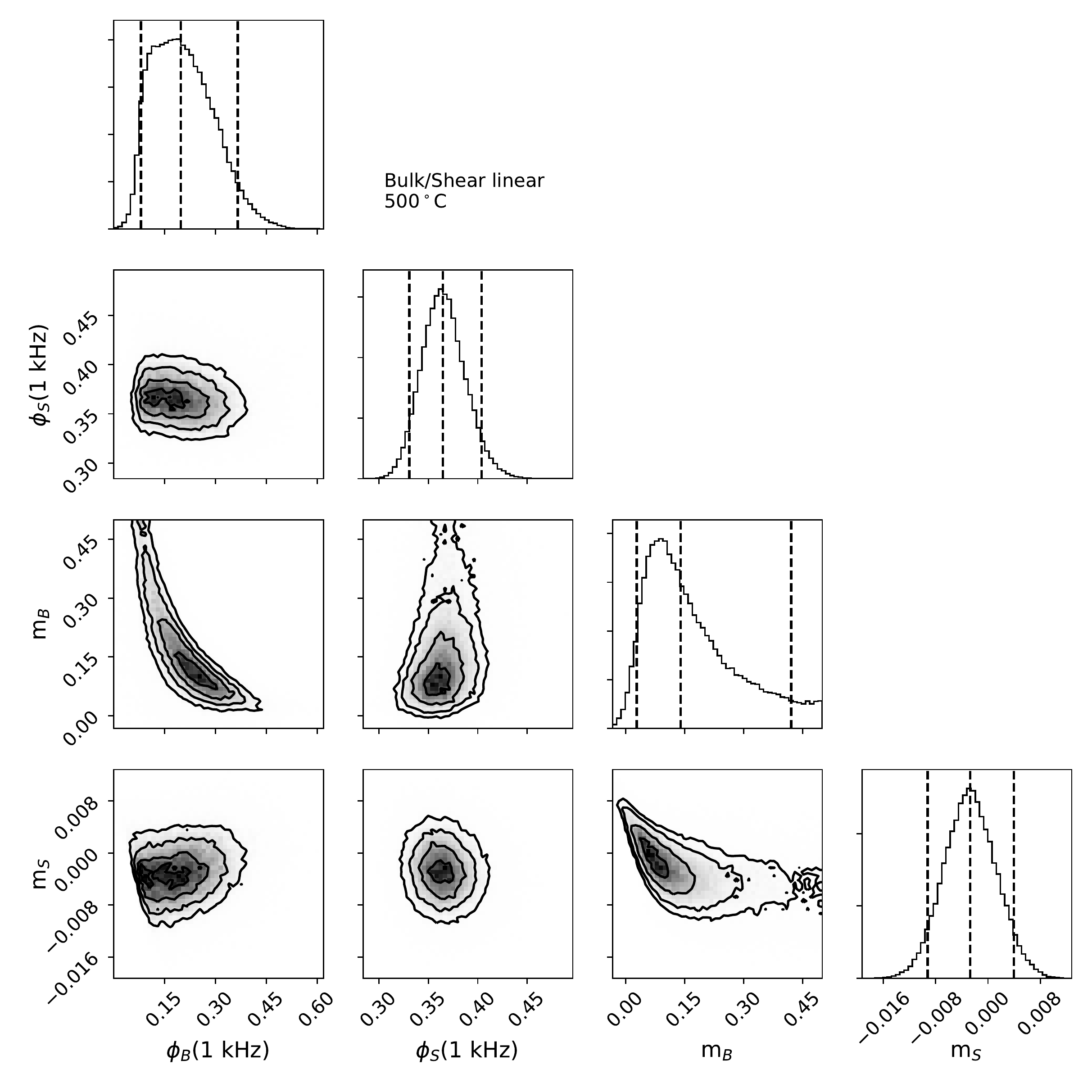} 
\end{center}
\caption{\label{fig:corner} Posterior probability distributions of the parameters of two loss models (left, one loss angle with linear frequency dependency; right, bulk and shear different loss angles with linear frequency dependency. The results shown here as an example, correspond to the measurements of titania-doped tantala films after annealing at 500$^\circ$C. The posterior probability distributions have been marginalized over the Young's modulus, the Poisson ratio, the film thickness and density.}
\end{figure*}

\begin{figure*}[tb] 
\begin{center}
\includegraphics[width=\columnwidth]{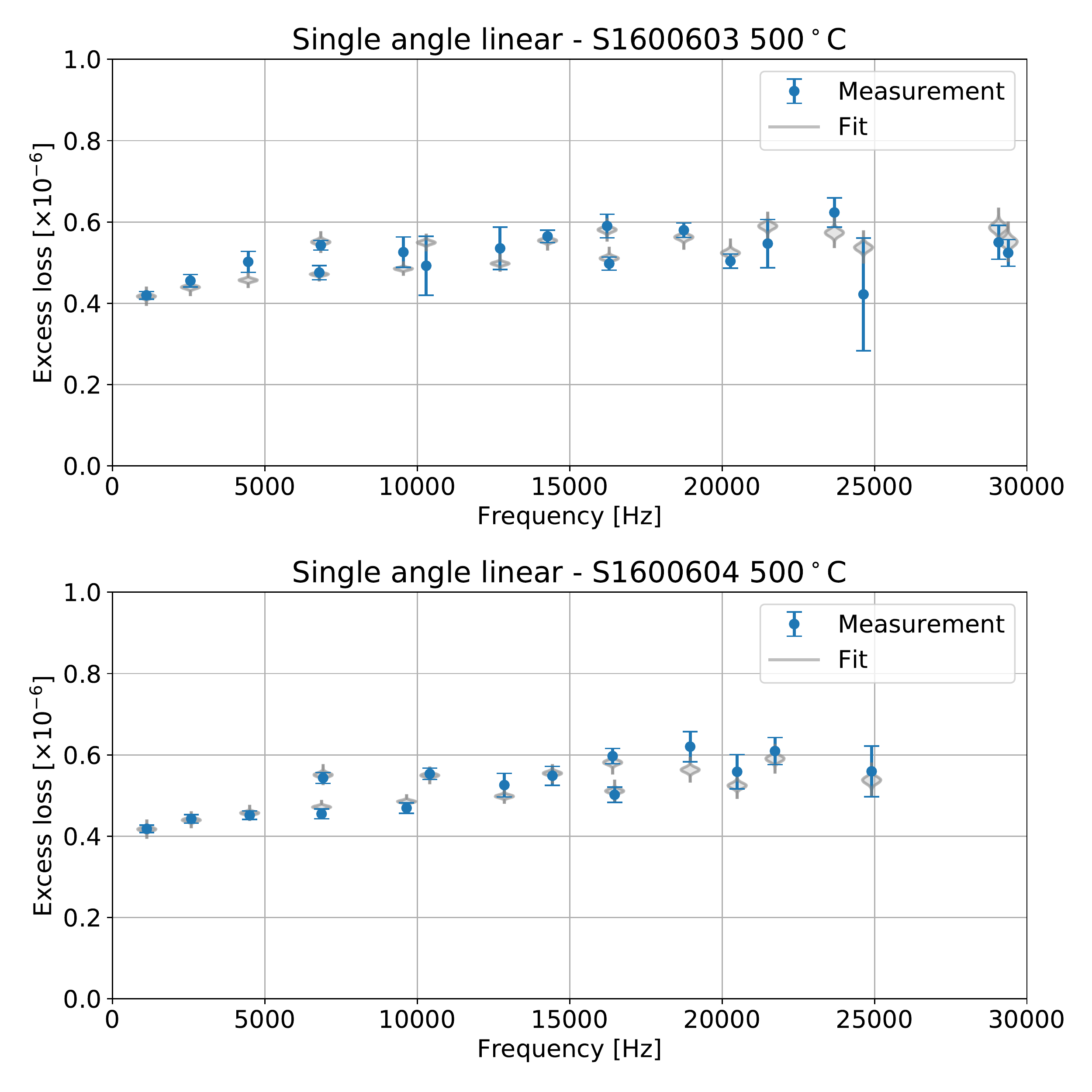} \includegraphics[width=\columnwidth]{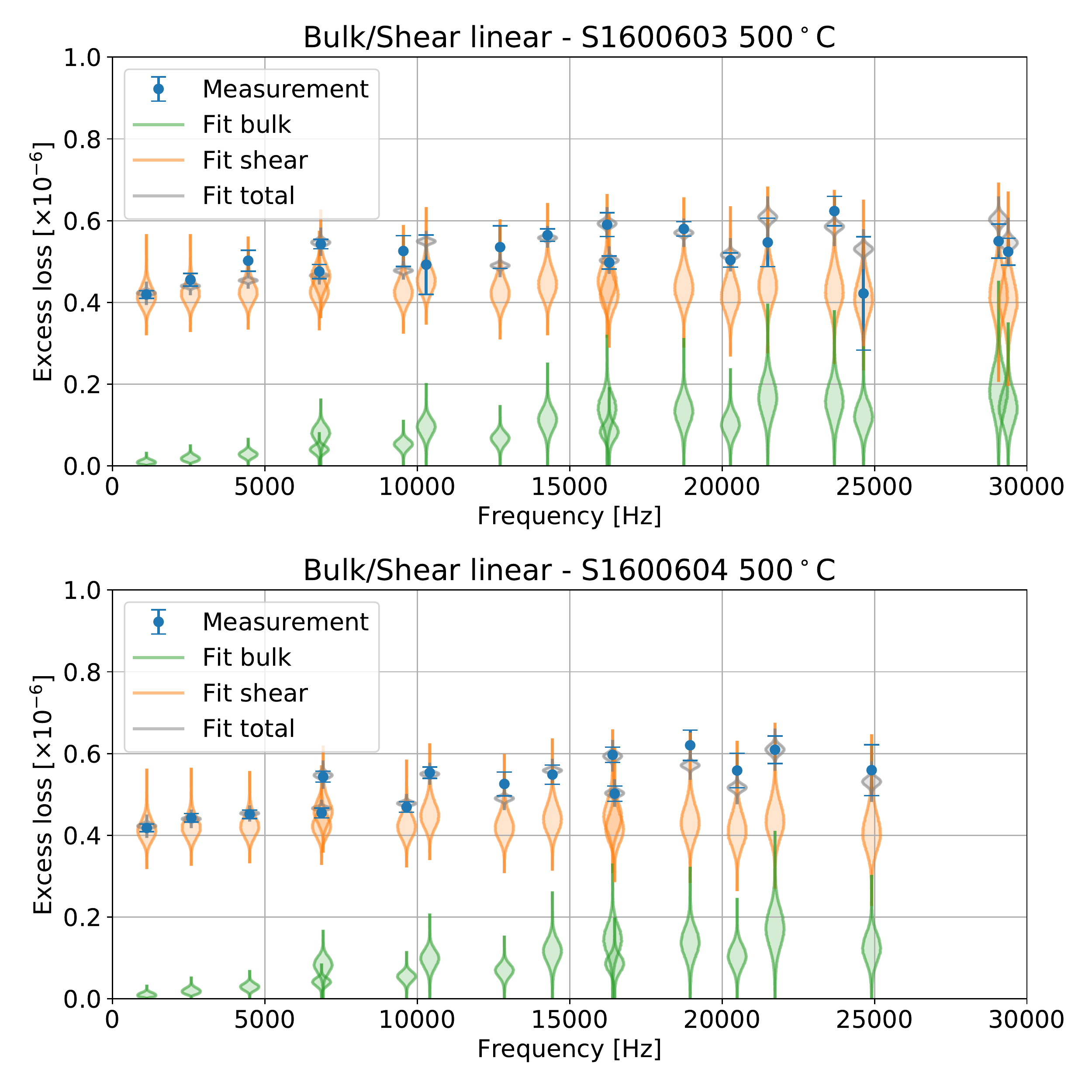} 
\end{center}
\caption{\label{fig:excess-loss} Comparison of the measured and predicted excess loss angle (not the material loss angle) for the two samples, named S1600603 and S1600604, and shown respectively in the top and bottom rows. The results shown here correspond to the samples measured after annealing at 500$^\circ$C. The left column shows in grey the distribution of the excess loss angle for the single loss angle model. The right column instead shows the distributions for the bulk and shear loss angle model: in green the bulk contribution, in orange the shear contribution in grey the sum of the two. In both columns, the error bars data points represent the measured values. The violin plots instead represent the distribution of the predicted values, given the result of the Bayesian analysis. }
\end{figure*}


\begin{table*}[bt]
\begin{center}
\begin{tabular}{|c|c|c|c|c|}
\hline
\textbf{Model 1}                 & \textbf{Model 2}              & \textbf{As deposited} & \textbf{Annealed 500$^\circ$C} & \textbf{Annealed 600$^\circ$C}  \\
\hline
Model 1                 & Model 2              & As deposited & Annealed 500$^\circ$C & Annealed 600$^\circ$C \\
Single angle, power law & Bulk/Shear linear    &  -15.5       &   -6.2                &  -18.1\\
Single angle, linear    & Bulk/Shear linear    &   -7.4       &   -1.6                &  -10.1\\
Bulk/Shear power law    & Bulk/Shear linear    &   -0.6       &   -0.2                &   -1.8\\
Single angle, power law & Bulk/Shear power law &  -14.9       &   -6.1                &  -16.3\\
Single angle, linear    & Bulk/Shear power law &   -6.8       &   -1.4                &   -8.3\\
\hline
\end{tabular}
\end{center}
\caption{Bayesian odd ratios of the models considered in this analysis. Every table entry shows the logarithm of the bayesian ratio of Model 2 over Model 1. Negative values means that the data favors Model 2. The bulk-shear angle, linear-frequency dependency is favored for all annealing temperatures. \label{tab:odd-ratio}}
\end{table*}

\begin{table*}[bt]
\begin{center}
\begin{tabular}{|c|c|c|c|c|}
\hline
\textbf{Heat treatment} & \textbf{Bulk loss at 1 kHz} & \textbf{Bulk loss slope} & \textbf{Shear loss at 1 kHz} & \textbf{Shear loss slope} \\
& $\phi_{1,B}$ [$10^{-3}$] & $m_{B}$  & $\phi_{1,S}$ [$10^{-3}$]  & $m_{S}$ \\
\hline
 30$^\circ$C  & $0.19 \pm 0.15$  & $0.24 \pm 0.19$ & $0.72 \pm 0.07$ & $-0.005 \pm 0.004$ \\
500$^\circ$C & $0.20 \pm 0.14$ & $0.14 \pm 0.20$ & $0.37 \pm 0.04$ & $-0.003 \pm 0.007$ \\
600$^\circ$C & $0.31 \pm 0.11$ & $0.09 \pm 0.07$ & $0.26 \pm 0.03$ & $-0.012 \pm 0.007$ \\
 \hline
\end{tabular}
\end{center}
\caption{Parameters for the best fit to the data in terms of bulk and shear loss angles, with a linear dependency on frequency. The values quoted are the median of the probability distribution of each parameter given the data, and the 90\% confidence intervals. \label{tab:bulk-shear}}
\end{table*}

\begin{figure*}[tb] 
\begin{center}
\includegraphics[width=2\columnwidth]{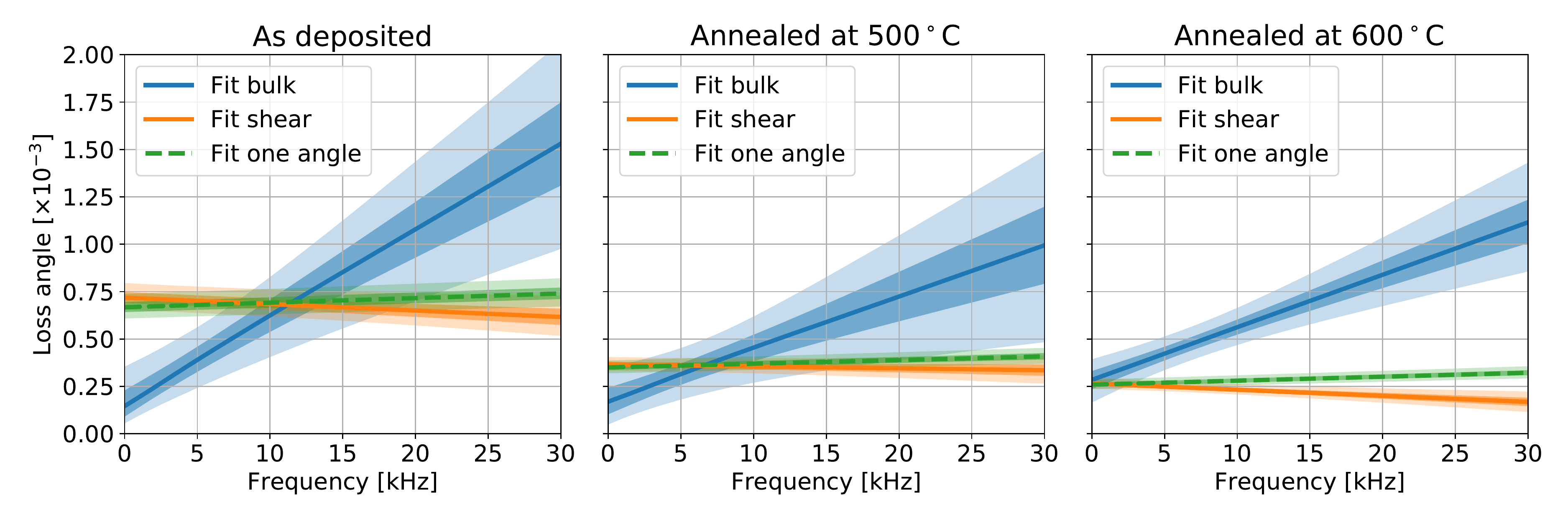} 
\end{center}
\caption{\label{fig:loss-angle} Estimated loss angles as a function of frequency for the measured titania-doped tantala film, after each heat treatment step. In each panel, blue and orange shows the best fit to bulk and shear loss angles respectively, while the green dashed line correspond to the best fit to a single loss angle model. }
\end{figure*}

The main goal of this work is to determine which material loss angle(s) model describes better the experimental data points. For each set of measurements (as deposited samples or annealed samples), we model the excess loss angle assuming either equal or different bulk and shear loss angles for the film material. For both model choices, we allow for a frequency dependency of the loss angles, in the form of a power law or a linear relationship:
\begin{eqnarray}
\phi_{\mathrm{power law}}(f; \phi_1, \alpha) &=& \phi_{1} \left( \frac{f}{1 \mbox{ kHz} }\right)^\alpha  \\
\phi_{\mathrm{linear}}(f; \phi_1, m) &=& \phi_1 \left( 1 + m \frac{f - 1 \mbox{ kHz}}{1 \mbox{ kHz}} \right)
\end{eqnarray}
where $\phi_1$ is the loss angle at 1 kHz, $\alpha$ is  the exponent of the power law, and $m$ the slope of the linear relationship. The excess loss angles measured experimentally are then modeled either with one loss angle, or with different bulk and shear loss angles:
\begin{eqnarray}
\delta \phi_i &=& D_i \phi_{x} (f_i; \phi_1, m) \\
\delta \phi_i &=& D_{B,i} \phi_{x}(f_i; \phi_{1,B}, m_B) \nonumber \\
&&+ D_{S,i} \phi_{x}(f_i; \phi_{1,S}, m_S) 
\end{eqnarray}
where $x$ can refer either to the linear or the power law relation, for a total of four different models that could describe the data: single loss angle with linear frequency dependency, single loss angle with power law frequency dependency, bulk and shear loss angles with linear frequency dependency, and bulk and shear loss angles with power law frequency dependency. To quantitatively determine which one of those four models better fits the measured data, we follow a Bayesian approach, which provides us with the probability distribution of the parameters for each model, and also the relative probability of the models, given the measured data set. In this section we briefly outline the basics of the Bayesian approach, with particular emphasis to its application to the problem at hand. The reader unfamiliar with Bayesian inference analysis should refer, for example, to \cite{bayesian, silvia2006, bayesian2011} for a more detailed description.

For each model, we want to compute the probability distribution $\mathcal{P}(\theta | \mathcal{M}_j, \delta\phi_i)$ of the parameters $\theta$ (for example $\left\{ \phi_1, \alpha \right\}$ in the case of the single loss angle, power law model) given the measured data $\left\{ \delta\phi_i\right\}$ and assuming one of the models, $\mathcal{M}_j$, to be valid. This probability distribution is usually called the \emph{posterior distribution} of the model parameters. To compute it, we use Bayes' theorem \cite{bayesian}:
\begin{equation}\label{eq:bayes}
\mathcal{P}(\theta | \mathcal{M}_j, \delta \phi_i) = \frac{\mathcal{P}(\delta\phi_i | \mathcal{M}_j, \theta) \cdot \mathcal{P}(\theta | \mathcal{M}_j)}{\mathcal{P}(\delta \phi_i | \mathcal{M}_j)}
\end{equation}
where the term $\mathcal{P}(\delta\phi_i | \mathcal{M}_j, \theta)$ describes the probability (\emph{likelihood}) of obtaining the measured data given the model and a specific value of the parameters, and the term $\mathcal{P}(\theta | \mathcal{M}_j)$, usually called the \emph{prior} probability distribution of the parameters, encodes our knowledge of the possible values of the parameters, given a specific model, before any measurement is taken. Finally, the term at the denominator $\mathcal{P}(\delta \phi_i | \mathcal{M}_j)$ is the probability of obtaining the measured data if the model is assumed, and allowing any value for the parameter. This last term can be computed as a normalization, by integrating the left hand side of equation \ref{eq:bayes} over all values of $\theta$ and requiring the result to be equal to one, since it is a probability distribution. This term will play a role in the later selection of the most likely model.

In our case, the data consist of the measured excess loss angle $\delta \phi_i$ for both the samples measured, for each of the accessible resonant mode frequencies, with the measurement uncertainties. For any of the models, the data likelihood $\mathcal{P}(\delta\phi_i | \mathcal{M}_j, \theta)$ is modeled as a normal distribution, where each data point is an independent random variable with variance given by the experimental uncertainties in the measured quality factors. For each model, the parameter set $\theta$ is composed of two parts. First, we allow the film properties to vary within the uncertainties described in section \ref{sec:measurements}: the Young's modulus and Poisson ratio have normal probability distributions centered on the best fit of the resonant mode frequency shifts, with variance given also by the fit, as reported in table \ref{tab:parameters}; the coating density and thickness are also allowed to vary with a normal probability distribution centered on the nominal value and with a variance corresponding to a $5$\% uncertainty as explained in section \ref{sec:measurements}. Secondly, the prior distributions of the other model parameters are assumed to be flat: the loss angle at 1 kHz can vary in the range $\phi_1 \in [0, 3 \times 10^{-3}]$ for all models; for the power law loss angle models the exponent can vary in the range $m \in [-2,2]$, while for the linear models the slope is restricted to values that exclude negative loss angles $m \in [-0.033, 0.5]$. As we shall see, the results are not very sensitive to the choice for the allowed range of the parameters, meaning that the measured data is increasing our knowledge of the models, as expected.

There are many ways to use equation \ref{eq:bayes} to compute the posterior distribution of the model parameters. The method most commonly used, and also adopted for this work, is to numerically sample the posterior distribution, or in other words to compute a large set of points in the parameter space, distributed in a way that follows the posterior distribution. We carried out this sampling using a Markov Chain Montecarlo (MCMC) algorithm implemented with the Python package \texttt{emcee} \cite{emcee}. The results can then be used to numerically evaluate the distribution of each parameter. Since the model parameter space is high dimensional, it is impossible to represent graphically the full distribution. We therefore plot the sets of all joint  distributions of pairs of parameters. The results are shown in figure \ref{fig:corner} for the two samples annealed at 500$^\circ$C, and considering the following two models: one single loss angle with linear frequency dependency, or bulk and shear different loss angles with linear frequency dependencies (similar results are available for all annealing temperatures and the power law models, but they are not shown here for brevity). Each panel in the two \emph{corner plots} show the joint probability distribution for pairs of parameters, as well as the probability distribution of each parameter, at the top of each column. Each of the contour plots in figure \ref{fig:corner} represents the probability distribution of the two parameters, given the data and assuming one of the models. All the other parameters are allowed to take any value, a procedure often referred as \emph{marginalization}. The one-dimensional histograms show the probability distribution of each parameter, marginalized over all the others. The dashed lines represent the 90\% confidence intervals and the median of the posterior distributions. Those values can be taken as the best estimates and uncertainties of the parameters, given the data and assuming one specific model.

Once the posterior distribution of all model parameters is so obtained, we can compute the distribution of the excess loss angle for each resonant mode and compare the results with the experimental measurements. This is done by using each point in the parameter space obtained from the MCMC sampler in the corresponding model to compute the excess loss, and then producing a histogram of all values. Figure \ref{fig:excess-loss} shows the results for both model considered here as an example: single loss angle with linear frequency dependency and different bulk shear loss angles, again with linear frequency dependency (similar results for all annealing temperatures and power law frequency dependency are also available, but not shown here for brevity). In those plots the distribution of the excess loss angles are shown and compared with the experimental results. In the case of the bulk and shear loss angle model, both contributions are shown separately, together with the sum. One can notice that most of the excess loss angle is due to the shear contribution, but there is nevertheless a not negligible contribution coming from the bulk losses.

The Bayesian approach we used to fit the model parameters allows us to compute the probability of the different models $\mathcal{P}(\mathcal{M}_j | \delta \phi_i)$, given the measured data points. Using Bayes' theorem again, this can be written as
\begin{equation}
\mathcal{P}(\mathcal{M}_j | \delta \phi_i) = \frac{\mathcal{P}(\delta \phi_i | \mathcal{M}_j) \mathcal{P}(\mathcal{M}_j)}{\mathcal{P}(\delta \phi_i)}
\end{equation}
where $\mathcal{P}(\mathcal{M}_j)$ is the prior probability of the models, and $\mathcal{P}(\delta \phi_i | \mathcal{M}_j)$ is the likelihood of obtaining the measured data points given the model. The latter can be computed from the results of the MCMC sampler as explained above. The term in the denominator acts as a normalization constant, independent of the model. Therefore, assuming all models are equally likely a priori, we can compute the logarithm of the Bayesian odd ratio of any pair of models, given the data: 
\begin{equation}
\log O(\mathcal{M}_1, \mathcal{M}_2) = \log \left[ \frac{P(\mathcal{M}_1 | \delta \phi_i)}{P(\mathcal{M}_2 | \delta \phi_i)} \right]
\end{equation}
A logarithm odd ratio greater than zero means that the measured data favors the model at the numerator $\mathcal{M}_1$, while a value lower than zero means that the model at the denominator $\mathcal{M}_2$ is favored. We use the Bayesian odd ratios to determine which model is favored by the data, since this approach takes naturally into account the uncertainty in the data points and in the estimated film mechanical properties, as well as the different dimensionality of the parameter space for each model. It also provides a quantitative measurement of the "goodness of the fit" based on the model complexity and measurement uncertainties.

Table \ref{tab:odd-ratio} lists the logarithm of the odd ratio for pairs of models. For all the annealing temperature, as well as for the as deposited film, the measured data strongly favor the models with different bulk and shear loss angles. Among those models, the linear frequency dependency is slightly favored. Table \ref{tab:bulk-shear} summarizes the estimated parameters for this model. Figure \ref{fig:loss-angle} shows the results in graphical form. In the same plot we compare the bulk and shear loss angles with the estimate obtained using a single loss angle model, as done in most of previous work. 

\begin{figure}[tb] 
\begin{center}
\includegraphics[width=\columnwidth]{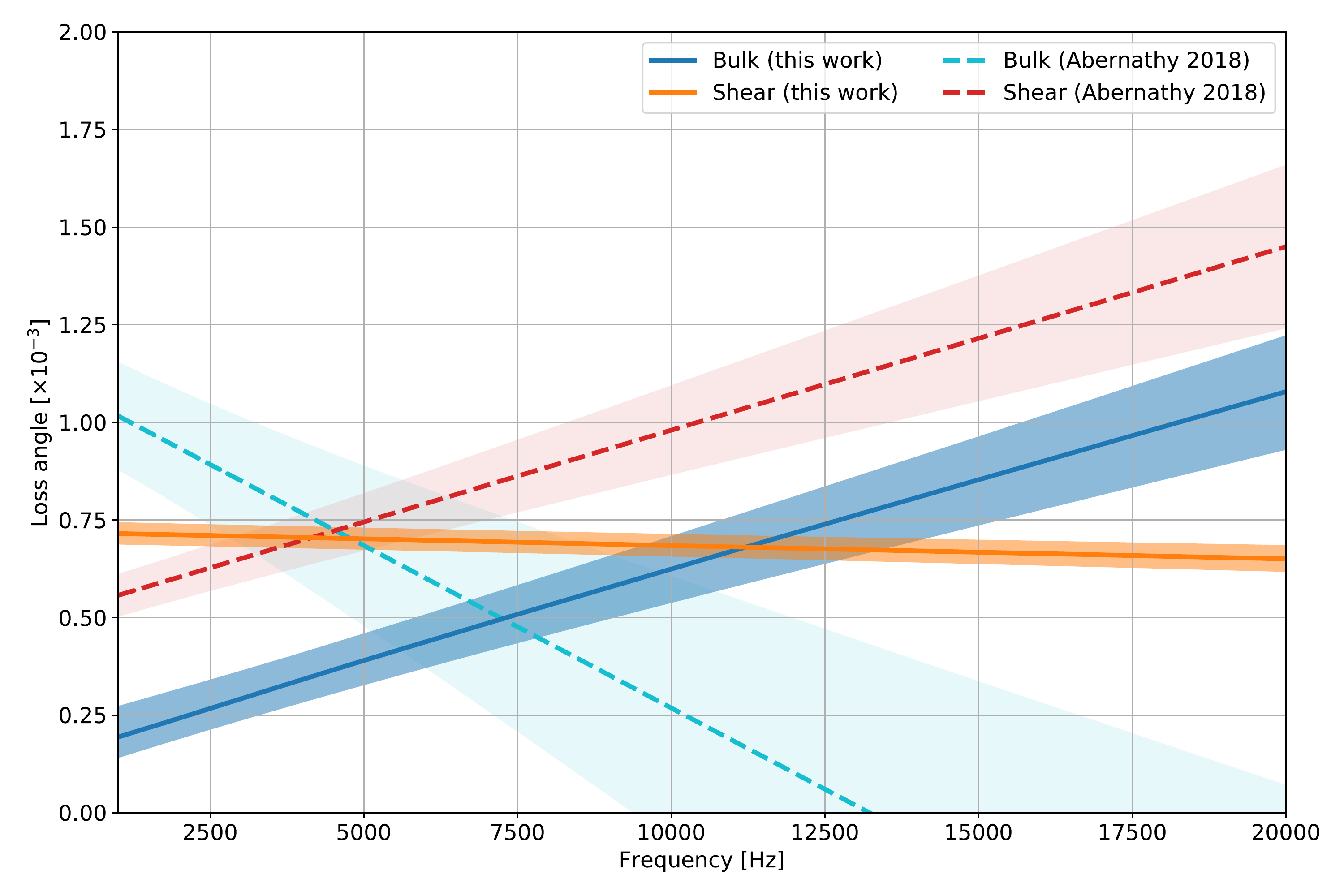} 
\end{center}
\caption{\label{fig:abernathy} Comparison of bulk and shear loss angles for the as deposited titania-doped tantala, as obtained in this work and as reported in Abernathy et al. \cite{abernathy2018}.}
\end{figure}

Figure \ref{fig:abernathy} compares our results for the as-deposited film with those reported in Abernathy et al. \cite{abernathy2018}, where a similar analysis was performed. Our results are not consistent with those reported in that work, showing opposite frequency dependencies and different relative amplitude of the two loss angles.  We should note that the two films, although both being made of about 20\% titania doped tantala, were produced by different groups employing different coating deposition chambers (in our case, films were grown by reactive ion beam sputtering using the Laboratory Alloy and Nanolayer System manufactured by 4Wave, Inc \cite{lans} at Colorado State University; in Abernathy's case, an ion beam sputtering system was used by the Commonwealth Scientific and Industrial Research Organization \cite{CSIRO}) and therefore might have different properties. If we assume that the two films have similar properties, the reason for the discrepancy is not understood at the moment of writing. However, we would like to point out some key differences between the measurement reported in Abernathy et al. \cite{abernathy2018} and our results: the samples were suspended with different techniques, which might induce systematic differences; we measured and subtracted the contribution to the loss angle of the uncoated substrate, while it is not clear how that was treated in Abernathy's work; in our work a larger number of resonant mode was probed; in Abernathy's work bulk and shear loss angles are extracted from pairs of Q measurement, assuming no frequency dependency between the two modes in each pair but allowing for a frequency dependency between pairs, while in our work we directly fit a frequency dependent model to the experimental data; finally, in our work we restricted the fit parameters to physically realizable values, while in Abernathy's the bulk loss angle is predicted to have negative values for high frequencies.

In this analysis the film is assumed to have uniform thickness and mechanical properties, and to cover the entire substrate surface. The expected variation of the film thickness over the surface is expected to be small. However, variations of the film properties with position might introduce mode-dependent systematic errors that have not been considered in this study. Further work is needed to quantify their effect on the bulk and shear loss angle results.

In previous works \cite{cagnoli2018}, the mechanical quality factors of uncoated silica disks were found to be dependent and limited by loss mechanisms at the unpolished edge, and were also found to degrade over time. The silica disks used in this work have an optical quality polished edge, and the mechanical quality factors have been measured before the film deposition, to ensure a correct subtraction of the background due to the substrate. We also verified that the polished edge ensures that there is no significant evolution of the substrate quality factor over time. Therefore we are confident that the effect described in \cite{cagnoli2018} is not an issue in our work.

\section{Effect on thermal noise estimate}\label{sec:thermal-noise}

\begin{figure}[tb] 
\begin{center}
\includegraphics[width=\columnwidth]{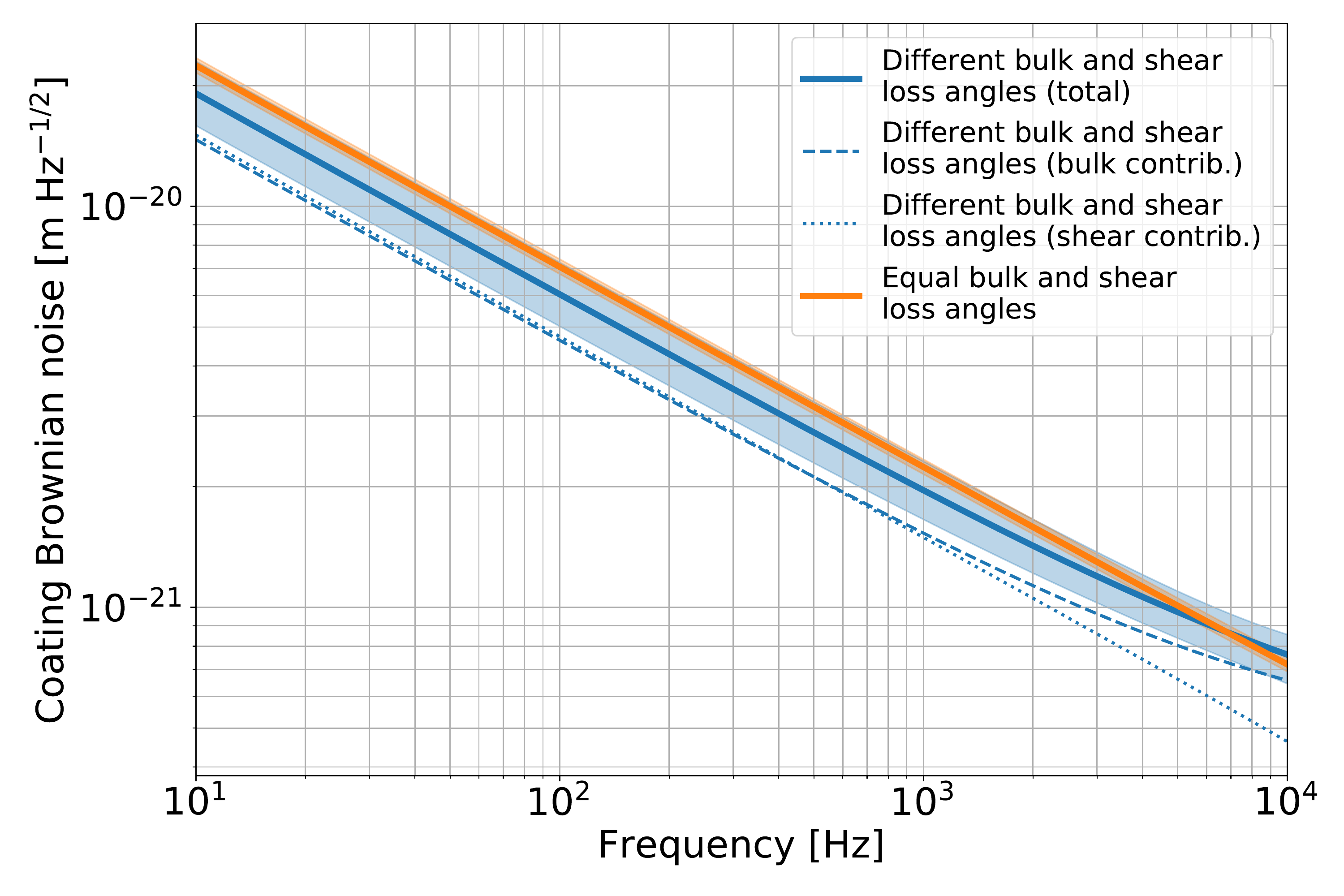} 
\end{center}
\caption{\label{fig:thermal-noise} Brownian noise for a single high reflectivity mirror, composed of alternating layers of silica and titania-doped tantala, as described in the main text. The solid orange line shows the displacement noise using the model where the bulk and shear loss angles are equal, while the solid blue line corresponds to the model where bulk and shear can assume different values. The dashed and dotted curves show the bulk and shear contribution, respectively. }
\end{figure}

The standard computations used to estimate the contribution of coating thermal noise in the advanced gravitational wave detectors \cite{thermal-noise} assume that both the low and high index materials can be described with one single loss angle. Direct thermal noise measurements have also been performed \cite{mit} and the results expressed again in terms of equal bulk and shear loss angles. Here we use the result of our analysis, and compute the expected thermal noise for a high reflectivity mirror similar to the design employed in the Advanced LIGO detectors, using the inferred bulk and shear loss angles. We use the model described in Hong et al. \cite{hong2013} (in particular starting from equation 94 therein), where the properties of the component materials and the geometry of the layers are used to predict the total thermal noise. Possible effects due to the transition between layers are not considered \cite{prd93, coatings-gw}.

We consider a high reflection coating composed of 38 alternating layers of silica (low index material) and titania-doped tantala (high index material), each with an optical thickness of $\lambda/4$ where the laser wave-length $\lambda$ in vacuum is $1064$ nm, to obtain a nominal transmission of about 5 ppm \cite{amato2018}. For the titania-doped tantala loss angle we use the results reported in this work, for the film measured after annealing at 500$^\circ$C. We compare two different cases: the best fit to a single loss angle and the best fit with different bulk and shear loss angles, as shown in figure \ref{fig:loss-angle}. The contribution of silica to thermal noise is small, but nevertheless we included a frequency dependent model obtained from another measurement we performed on silica thin films annealed at 500$^\circ$C. In this case the sensitivity of our ring-down measurement was not enough to disentangle bulk and shear loss angles: the experimental data is best described by a single loss angle, linearly dependent on the frequency, given by 
\begin{eqnarray*}
\phi_{\mathrm{SiO}_2}(f) &=& (0.035 \pm 0.004)\times 10^{-3} \cdot  \\
&& \left[ 1 + (-0.006 \pm 0.007)\times 10^{-3} \frac{f - 1 \mbox{ kHz}}{1 \mbox{ kHz}} \right]
\end{eqnarray*}

Figure \ref{fig:thermal-noise} shows the displacement noise due to the Brownian noise of a single high reflectivity mirror. As a reference, assuming the best fit to the data with a single loss angle, we obtain a coating Brownian noise of $(7.0 \pm 0.3) \times 10^{-21}$ m/$\sqrt{\mbox{Hz}}$ at 100 Hz. Using instead the best fit to the data with different bulk and shear loss angles, we obtain $(6.0 \pm 1.1) \times 10^{-21}$ m/$\sqrt{\mbox{Hz}}$ at 100 Hz. For comparison, the direct thermal noise measurement reported in \cite{mit} can be extrapolated to a level of $(7.5 \pm 0.1) \times 10^{-21}$ m/$\sqrt{\mbox{Hz}}$ at 100 Hz. Within the precision of our measurement, there is no significant impact on the estimate of thermal noise for and Advanced-LIGO-like high reflectivity coating.

It is worth noting that the knowledge of the separate bulk and shear loss angles could allow an additional degree of freedom to optimize the thermal noise of the coating, by changing the thickness of the layers \cite{hong2013}. 

\section{Conclusions}\label{sec:conclusions}

We showed that it is possible to estimate the bulk and shear contribution to the loss angle of a thin film, using measurements of the decay time of the resonant modes of a coated silica disk, carried out in a Gentle Nodal Suspension system. As an example we analyzed a thin film of titania-doped tantala, one of the materials used in the advanced gravitational wave interferometric detector mirrors. A Bayesian analysis of the experimental data shows that a model featuring different bulk and shear loss angle is favored with respect to a simpler model with one single loss angle (i.e. same loss angle for bulk and shear energies). The change in loss angles with annealing is more evident in the shear  than in the bulk contribution. When the two models are used to compute the expected thermal noise for a high reflection mirror similar to those used in Advanced LIGO, the difference is marginal and within error bars when the measurements are extrapolated in the frequency region between 10 and 1000 Hz.

\section*{Acknowledgments}
LIGO was constructed by the California Institute of Technology and Massachusetts Institute of Technology with funding from the United States National Science Foundation under grant PHY-0757058. This work was also supported by the Center for Coating Research, NSF grants PHY-1708010 PHY-1710957 and by the LIGO Laboratory, NSF grant PHY-1764464. The authors are grateful for computational resources provided by the LIGO Laboratory and supported by the National Science Foundation Grants PHY-0757058 and PHY-0823459. This paper has LIGO document number P1900336. 


\end{document}